\documentclass[10pt,twocolumn,aps,pre,superscriptaddress]{revtex4-1}

\usepackage{amsmath}
\usepackage{graphicx}
\usepackage{color}
\usepackage{upgreek}
\usepackage[linkcolor = blue, citecolor = blue, urlcolor = blue, colorlinks = true]{hyperref}
\usepackage[version=3]{mhchem}
\usepackage{commath,amssymb}
\usepackage{ushort}
\usepackage[dvipsnames]{xcolor}
\usepackage{tikz}
\usepackage{amssymb}
\usepackage{xfrac}

\begin{document}
\author{J. de Graaf}
 \email{j.degraaf@uu.nl}
 \affiliation{Institute for Theoretical Physics, Center for Extreme Matter and Emergent Phenomena, Utrecht University, Princetonplein 5, 3584 CC Utrecht, The Netherlands}

\author{W. C. K. Poon}
 \affiliation{SUPA, School of Physics and Astronomy, The University of Edinburgh, King's Buildings, Peter Guthrie Tait Road, Edinburgh, EH9 3FD, United Kingdom}

\author{M. J. Haughey}
 \affiliation{SUPA, School of Physics and Astronomy, The University of Edinburgh, King's Buildings, Peter Guthrie Tait Road, Edinburgh, EH9 3FD, United Kingdom}

\author{M. Hermes}
 \affiliation{Soft Condensed Matter, Debye Institute for Nanomaterials Science, Utrecht University, Princetonplein 5, 3584 CC Utrecht, The Netherlands}


\title{Hydrodynamics strongly affect the dynamics of colloidal gelation but not gel structure}
\date{\today}

\begin{abstract}
Colloidal particles with strong, short-ranged attractions can form a gel. We simulate this process without and with hydrodynamic interactions (HI), using the lattice-Boltzmann method to account for presence of a thermalized solvent. We show that HI speed up and slow down gelation at low and high volume fractions, respectively. The transition between these two regimes is linked to the existence of a percolating cluster shortly after quenching the system. However, when we compare gels at matched `structural age', we find nearly indistinguishable structures with and without HI. Our result explains longstanding, unresolved conflicts in the literature.
\end{abstract}

\maketitle

\section{\label{sec:intro}Introduction}

Many applications require particulates to remain dispersed throughout a liquid medium against their tendency to sediment or cream due to density mismatch, often aggravated by a tendency of the suspension to phase separate. Thus,~\textit{e.g.}, sedimentation in a drug suspension can have dangerous consequences for patients. For colloids with short-ranged attractions, gravitational stability can be achieved by quenching the system to form a space-spanning network, a colloidal gel. In many systems, gelation is due to the arrest of this phase separation process by the glass transition~\cite{manley2005glasslike, zaccarelli2007colloidal}.

Kinetics do not determine the equilibrium structure of a system. However, in colloidal gels, which are arrested states, the kinetics of aggregation crucially affect structure formation and therefore mechanical properties~\cite{koumakis2015tuning}. Hydrodynamic interactions (HI) clearly affect kinetics and hence may lead to a lack of predictive power for numerical simulations of gelation that do not account for these. Simulating a system with HI is, however, far from trivial, as historically this has been the most computationally expensive part of such a simulation algorithm~\cite{batchelor1972determination, brady1988stokesian}. Concerted effort over the last two decades comparing simulations with and without HI has nevertheless suggested a plethora of effects.

The earliest forays into this topic considered the clustering of a few particles into ``fractal'' aggregates. Tanaka and Araki~\cite{tanaka2000simulation} showed that without HI, two-dimensional (2D) Lennard-Jones-like (LJ-like) particle aggregation leads to more compact clusters. Whitmer and Luijten~\cite{whitmer2011influence} extended this to three-dimensional (3D) clustering for particles interacting via an Asakura-Oosawa-like (AO-like) potential, finding the diffusion-limited cluster aggregation (DLCA) result at high interaction strength~\cite{witten1981diffusion}.

Turning from clusters to gels, Yamamoto~\textit{et al.}~clarified the hydrodynamic stabilization of 2D LJ-type gels formed by quenching to zero temperature~\cite{yamamoto2008role}. However, in 3D (colloid volume fractions $\phi = 0.173$ and $0.307$) they found only a minor quantitative effect of HI, observing that structural evolution does not cease in 3D as it does in 2D~\cite{yamamoto2008role}. Furakawa and Tanaka~\cite{furukawa2010key} investigated the time evolution of 3D AO-like gels with $\phi = 0.13$ and $\Delta U = 4.2 k_{\mathrm{B}}T$ and $8.4 k_{\mathrm{B}}T$ quenches, with $\Delta U$ the contact attraction strength, $k_{\mathrm{B}}$ Boltzmann's constant, and $T$ the temperature. HI shifted the gelation line and changed the time dependence of the correlation length in the gel, $\xi \sim t^{x}$, from $x = \sfrac{1}{3}$ to $\sfrac{1}{2}$.

Recently, Varga and collaborators performed large-scale simulations of colloidal gelation using a highly accelerated hydrodynamics algorithm~\cite{swan2016rapid}. They contrasted systems with and without HI over a wide range of $\phi$ and $\Delta U$, with both AO-like potentials~\cite{varga2015hydrodynamics} and a combination of short-ranged attraction and long-ranged repulsion~\cite{varga2016hydrodynamic}, finding that HI could lower the gelation line~\cite{varga2015hydrodynamics}, slow down compaction and speed up coagulation~\cite{varga2016hydrodynamic}. Royall~\textit{et al.} reported a similar lowering of the gelation line with the inclusion of hydrodynamic interactions, which they related to a change in the local structure of colloidal gels with the inclusion of HI~\cite{royall2015probing}.

In this work, we consider gelation with and without HI for colloids with short-ranged attractions ($\Delta U = 5 k_{\mathrm{B}}T$ and $10 k_{\mathrm{B}}T$) and $0.075 \leq \phi \leq 0.225$. HI are accounted for using the lattice-Boltzmann (LB) method~\cite{dunweg2009lattice}, which enables us to simulate systems comparable or even larger than those of Varga~\textit{et al.}~\cite{varga2015hydrodynamics, varga2016hydrodynamic}. We use the void volume (VV)~\cite{haw2006void, koumakis2015tuning} to study the structure of the holes in the gel. This allows precise analysis of the dynamics of gelation and identification of the onset of aging, which agrees with the one obtained from the evolution of the correlation length. We find that HI indeed speeds up gelation and slightly steepens the VV power-law scaling for low $\phi$. For $\phi \gtrsim 0.16$, however, systems with HI exhibit the same power-law exponent as their non-HI counterparts and gelation is slowed down. We relate this crossover to the presence of large clusters immediately upon quenching into the spinodal region of the phase diagram, which rearrange quickly to achieve percolation. We demonstrate clearly that aging is, however, not related to percolation and typically sets in at a much later time.

We also study the structure of gels formed with and without HI. However, in contrast with most simulations to date, we do not compare the structure at equal-and-fixed time. Instead, we compare gels at equivalent points during their evolution, in which case, surprisingly, gels formed with and without HI cannot be readily distinguished by their structure factor. More sensitive VV analysis reveals a small influence of HI on structure, but the difference is much smaller than what can be achieved by minor variations in the gel age, $\phi$, or $\Delta U$.

Note that we focus throughout on $\phi \geq 0.075$. At lower $\phi$, gel formation is so slow that other effects such as gravity and convection will dominate the dynamics of any real system. Moreover, our method does not give enough statistics over long enough time to determine the effect of HI on the position of the gelation boundary. Nonetheless, our results suffice to show that HI have a large effect on the speed at which the quenched system evolves, but the path that it takes through state space is nearly unaffected. We discuss how these findings relate to and reconcile existing literature on HI in colloidal gelation.

\section{\label{sec:methods}Simulation Methods}

We evaluate the influence of HI on gelation by comparing simulations using strongly damped Langevin dynamics~\cite{hinch1975application} (no HI: NH) to those using a GPU-accelerated fluctuating LB method~\cite{rohm2012lattice} (with HI: WH). Particles are coupled to our LB solvent through the Ahlrichs and D{\"u}nweg method~\cite{ahlrichs1999simulation}, similar to a recent study on colloidal crystallization~\cite{roehm2014hydrodynamic}. This captures the relevant far-field hydrodynamics accurately~\cite{ahlrichs1999simulation} and approximates near-field effects. The lubrication regime is not well-described using the Ahlrichs and D{\"u}nweg method, but this regime is known not to influence gelation~\cite{bybee2009hydrodynamic}. Nevertheless, our particle contacts are always ``lubricated'', that is, we do not account for effects as mechanical friction that have recently come into prominence in aspects of suspension dynamics~\cite{lin2015hydrodynamic}. Lastly, it should be noted that in a real fluid two particles slow down when they approach each other because of hydrodynamic coupling; fluid must be squeezed out of the gap. This near-field effect, which already occurs outside of the lubrication regime, is qualitatively captured by the Ahlrichs and D{\"u}nweg method~\cite{rohm2012lattice} and important for the behavior of our system, as we will return to shortly.

We match the bulk diffusivity of our particles between the NH and WH simulations by setting $k_{\mathrm{B}}T = 1$, the particle diameter $\sigma = 1$, and the viscosity of our suspending medium to $\eta = 10$, so that $D = k_{\mathrm{B}}T / ( 3 \pi \eta \sigma ) = 1.1\times10^{-2} \sigma^{2} \tau^{-1}$, with unit time $\tau \equiv \sqrt{m \sigma^{2} / (k_{\mathrm{B}}T)}$ and particle mass $m = 1$. We express all times in terms of the Brownian time $t_{\mathrm{B}} = \sigma^{2}/(4 D) = 23.5 \tau$ for a particle to diffuse its own radius.

A generalized LJ interaction with high exponents was used to model short-ranged depletion attraction \cite{poon1995gelation}:
\begin{align}
\label{eq:LJgen} U_{\mathrm{LJ}}^{\mathrm{gen}}(r) &
\begin{cases}
= &\epsilon \left[ \left( \dfrac{\sigma}{r} \right)^{96} - 2 \left( \dfrac{\sigma}{r} \right)^{48} + c \right] \text{\;\; $r < r_{c}$} \\
= & 0 \;\;\; \text{otherwise}
\end{cases} .
\end{align}
Here, $r$ is the center-to-center distance, $r_{c}$ the cut-off distance, $\epsilon$ the interaction strength, and $c$ the shift in the potential. This form avoids the sharp cusp that occurs by adding the AO potential to hard-core repulsion, thereby improving the stability of our numerical integration. The range of the attraction is $\lesssim 6\%$ of the hard core (judged by where $U$ reaches $\pm k_{\mathrm{B}}T$). Throughout, we deem particles as belonging to the same cluster if $r < 1.05\sigma$.

We equilibrate our systems using a purely repulsive potential: $\epsilon = 10 k_{\mathrm{B}}T$, $c = 1$, and $r_{c} = \sigma$. We then quench into the spinodal region by instantaneously switching to an attractive interaction: $\epsilon = 10 k_{\mathrm{B}}T$ or $5 k_{\mathrm{B}}T$, $c = 0$, and $r_{c} = 1.5\sigma$, so that $\Delta U = 10 k_{\mathrm{B}}T$ or $5 k_{\mathrm{B}}T$. The reduced second virial coefficient~\cite{noro2000extended} for $U_{\mathrm{LJ}}^{\mathrm{gen}}(r)$ is
\begin{align}
\label{eq:vir2} b_{2} &= 1 + 3 \int_{\sigma}^{\infty} \mathrm{d}r \, r^{2} \left[ 1 - \exp \left( - \frac{U_{\mathrm{LJ}}^{\mathrm{gen}}(r)}{k_{\mathrm{B}}T} \right) \right] ,
\end{align}
with $b_{2} \approx -5.2 \times 10^{2}$ and $-4.7$ for $\epsilon = 10 k_{\mathrm{B}}T$ and $5 k_{\mathrm{B}}T$, respectively.

We use a time step $\Delta t = 0.002$ and equilibrate each sample for $50 t_{\mathrm{B}}$ before quenching to simulate gelation and aging for another $550 t_{\mathrm{B}}$ for $\phi > 0.1$ and $5500 t_{\mathrm{B}}$ for $\phi \le 0.1$. We prepare our systems in a cubic box with edge $L = 48\sigma$ (with $\sigma =$~LB spacing) for 12 points in $\phi \in [0.075,0.225]$, so that we simulate between $N=$~15,841 and 47,523 particles. For each data point, we run $10$ independent simulations using the Molecular Dynamics package \textsf{ESPResSo}~\cite{limbach2006espresso, arnold2013espresso} to obtain sufficient statistics.

\section{\label{sec:result}Results}

\begin{figure}[t]
 \centering
 \includegraphics[width=7.5cm]{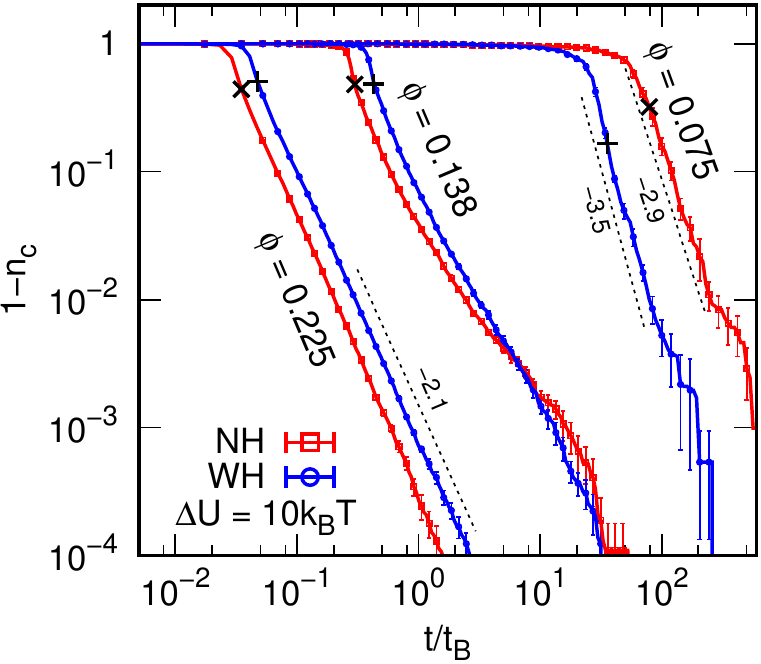}
 \caption{\label{fig:cluster}The fraction of particles in the largest cluster, $n_{\mathrm{c}}$, for $\Delta U = 10 k_{\mathrm{B}}T$ at three $\phi$ (labelled), as a function of time $t$ (in Brownian time units $t_{\mathrm{B}}$): \textcolor{red}{$\boldsymbol{\boxdot}$} = NH, \textcolor{blue}{$\boldsymbol{\odot}$} = WH; $\boldsymbol{\times}$ (WH) and $\boldsymbol{+}$ (NH) indicate percolation points; and error bars give standard errors. Black dotted lines and numbers indicate the exponent of asymptotic power laws, with standard error $\pm 0.1$.}
\end{figure}

\subsection{Percolation}

We start by examining the time evolution of the fraction of particles in the largest cluster $n_{\mathrm{c}}=N_\mathrm{c}/N$, Fig.~\ref{fig:cluster}, which shows the asymptotic approach to $n_{\mathrm{c}} = 1$ as a function of time, where we have indicated the time $t_{\mathrm{p}}$ at which each system first percolates ($\boldsymbol{+}$,$\boldsymbol{\times}$), which is at or immediately after the asymptotic behavior in $n_{\mathrm{c}}$ sets in. For low $\phi$ and $\Delta U = 10 k_{\mathrm{B}}T$ the asymptotic power law exponent is higher for WH than NH. This difference is less marked at $\Delta U = 5 k_{\mathrm{B}}T$ (data not shown), but the trend remains. For sufficiently high $\phi$, hydrodynamics slow down cluster growth. The power-law asymptotes are identical within errors, with little difference between $\Delta U = 10$ and $5k_{\mathrm{B}}T$. A crossover between these two kinds of behavior occurs at intermediate $\phi$ for both $\Delta U$.

At low enough $\phi$, small clusters form initially from particles already in close contact. These clusters diffuse, bonding on contact, leading to a fractal-like network. Such rearrangements are sped up by HI, which enhance cluster translational and rotational mobility by allowing collective effects, as found previously~\cite{varga2015hydrodynamics}. At higher $\phi$, particles travel only a short distance to bond. Now, compaction dominates,~\textit{i.e.}, the approach of colloids to form a bond contributes more to the formation of a gel network than motion of (small) clusters. This approach is slowed down by the need to squeeze fluid from between particles, again agreeing with recent work~\cite{varga2015hydrodynamics}.

\begin{figure*}[!htb]
 \centering
 \includegraphics[width=15.0cm]{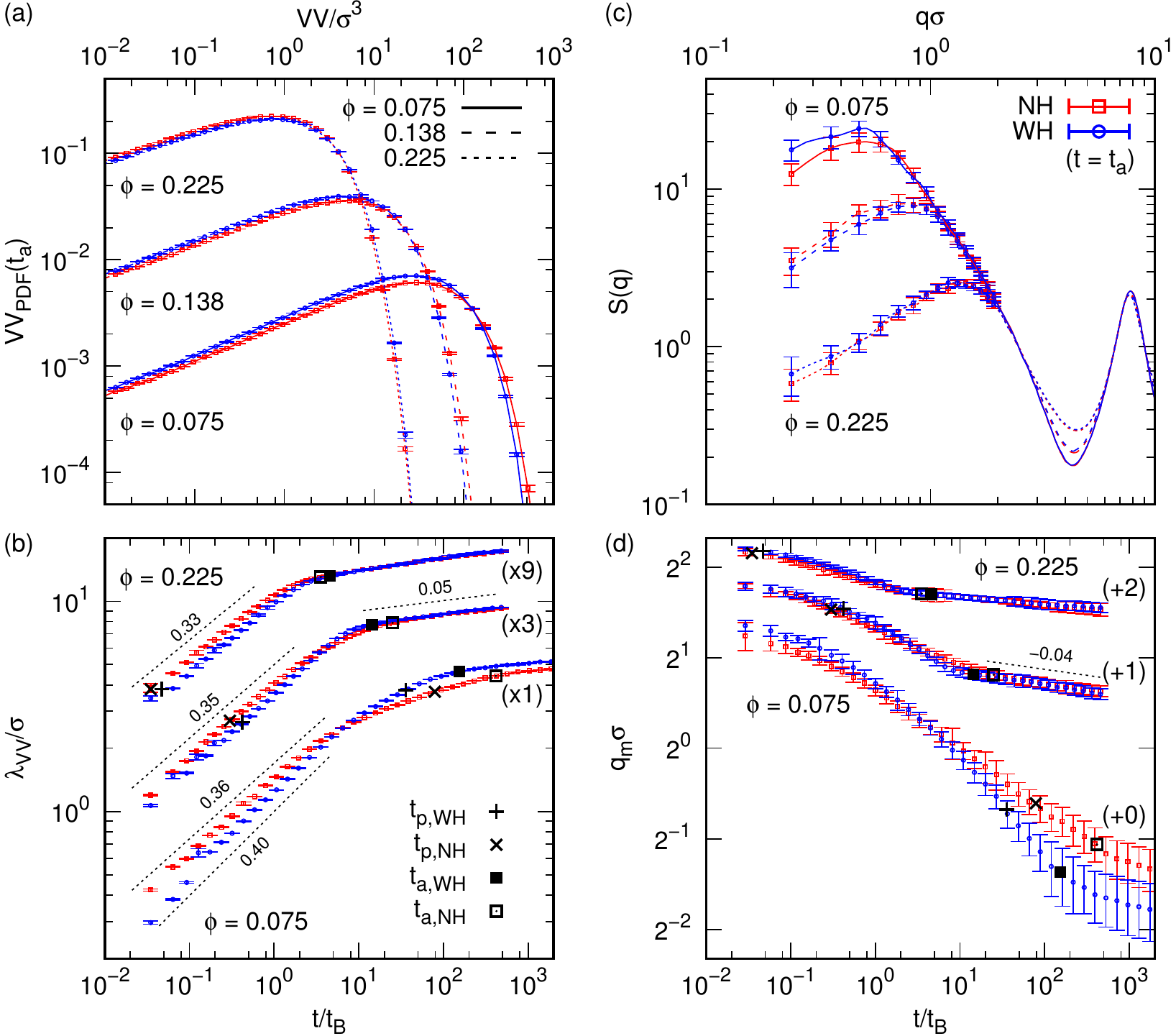}
 \caption{\label{fig:voids}Structural properties for gelling systems with $\Delta U = 10 k_{\mathrm{B}}T$, (\textcolor{red}{$\boldsymbol{\boxdot}$} = NH, \textcolor{blue}{$\boldsymbol{\odot}$} = WH), and for three $\phi$ as indicated by the labels and use of line type; the result for $\Delta U = 5 k_\mathrm{B}T$ is analogous. (a) Void volume (VV) probability density functions (PDFs) at the aging time $t_{\mathrm{a}}$. Error bars indicate the standard error, which is typically smaller than the symbol size. One in five data points is shown to improve the presentation, but the lines that serve as guides to the eye go through every data point. (b) The time dependence of the void growth length, $\lambda_{\mathrm{VV}}$, see Eq.~\eqref{eq:lambda}, again only one in five data points is shown. ($\boldsymbol{+}$,$\boldsymbol{\times}$) and ($\blacksquare$,$\square$) = percolation ($t_{\mathrm{p}}$) and aging onset times, respectively. The number in parentheses indicates the factor by which $\lambda_{\mathrm{VV}}$ is multiplied to separate the data. The thin dashed black lines and numbers indicate power-law fits and associated exponents (with error $\pm 0.005$), respectively. Only for $\phi = 0.075$ is there an appreciable difference in the NH/WH initial exponent. (c) Structure factors $S(q)$ at $t_{\mathrm{a}}$, corresponding to the data in (a). We only show error bars up to $q \sigma = 1$ for every third data point to improve the presentation, the curves connect all data points; the region of the central peak is not shown $q \sigma < 0.2$. (d) The peak position of the structure factor $q_{\mathrm{m}}$ as a function of time. The three WH/NH data sets have been shifted up by the amount indicated in parentheses to prevent overlaps. The dashed line indicates a power law with exponent $-0.04 \pm 0.005$.}
\end{figure*}

\subsection{Void-volume Analysis}

To identify structural differences between WH and NH systems, we calculate the void volume (VV). The VV of a point in space is the volume of a sphere centered at that point in contact with a particle surface~\cite{haw2006void, koumakis2015tuning}. Figure~\ref{fig:voids}a shows probability density functions (PDFs) for WH and NH systems at time $t = t_{\mathrm{a}}$ when aging starts, for $\Delta U = 10 k_{\mathrm{B}}T$ and three $\phi$. We will define aging once we have analyzed these VV PDFs to set the scene. 

Our low-noise PDF data over five orders of magnitude show small but systematic differences between the WH and NH cases. At $t_{\mathrm{a}}$, the WH systems at low $\phi$ have more small and fewer big holes than the NH systems, while the trend reverses for high $\phi$. The VV PDFs contain information that is complementary to the structure factor $S(q)$; we return to this shortly. 

From the PDFs, we compute the mean, $\langle \mathrm{VV} \rangle$, whose evolution gives a time-dependent length scale
\begin{align}
\label{eq:lambda} \lambda_{\mathrm{VV}}(t) &\equiv \left( \langle \mathrm{VV} \rangle (t) - \langle \mathrm{VV} \rangle(0) \right)^{1/3} ,
\end{align}
which measures the growth in void radius, Fig.~\ref{fig:voids}b. It better distinguishes the NH and WH systems and allows us to define aging.

There appears to be two power-law regimes $\lambda_{\mathrm{VV}} \propto t^{x}$. Initially, the WH ($x \approx 0.40$) and NH ($x \approx 0.36$) systems are slightly different at the lowest $\phi$, but for higher $\phi$ their power laws are identical within errors, $x \approx 0.35$ and $0.33$ respectively. This regime is associated with the gelation process. The second regime, which is less well developed at the lowest $\phi$, has exponent $0.05 \pm 0.005$, which we associate with aging.

We define the aging time, $t_{\mathrm{a}}$, as the crossover between these two regimes. In practice, this is done by fitting a local power-law $t^{x}$ to obtain a running exponent $x(t)$ and using the criterion $x = 0.08$ for the onset of aging, Fig.~\ref{fig:voids}b ($\blacksquare$,$\square$). The percolation times, $t_{\mathrm{p}}$, identified earlier ($\boldsymbol{+}$,$\boldsymbol{\times}$), are also shown. In each pair of data sets, we see that the NH and WH systems have the same $\lambda_{\mathrm{VV}}$ when they successively reach $t_{\mathrm{p}}$ and then $t_{\mathrm{a}}$. In other words, the gels have (nearly) the same structures when they have reached a comparable evolutionary stage. Observations for $\Delta U = 5 k_{\mathrm{B}}T$ are qualitatively similar, also reference Fig.~\ref{fig:voids}a.

\subsection{The Structure Factor}

The structure factor, $S(q)$, as a function of the wave vector $q$ at $t_{\mathrm{a}}$ for the WH and NH systems, Fig.~\ref{fig:voids}c, shows a broad, low-$q$ peak in all cases (at $q_{\mathrm{m}}$), defining a second characteristic length scale, $q_{\mathrm{m}}^{-1}$. As expected, $q_{\mathrm{m}}$ increases with $\phi$. Moreover, the $S(q)$ for WH and NH at $t = t_{\mathrm{a}}$ are identical within error bars at all $\phi$, supporting the conclusion we drew from the lower-noise VV data; Fig.~\ref{fig:voids}a. 

Fitting a time-dependent sequence of $S(q)$~\footnote{We obtained the location of the peak using simple sorting and binning, and fitted it to a local 4th order polynomial. Nevertheless, large errors remain due to the relatively large uncertainty in the measured $S(q)$.} yields $q_{\mathrm{m}}(t)$, Fig.~\ref{fig:voids}d, which can be directly compared to small-angle scattering experiments~\cite{poon1995gelation}, in which gelation is associated with the sudden slowdown in the decrease in $q_{\mathrm{m}}$ with time. The $t_{a}$ obtained using the VV analysis plotted in Fig.~\ref{fig:voids}d ($\blacksquare$,$\square$) matches this transition point well within errors, showing that it is the onset of aging rather than percolation ($\boldsymbol{+}$,$\boldsymbol{\times}$) that identifies gelation. The negative of the power law exponents obtained from our void-volume analysis, see Fig.~\ref{fig:voids}b, is consistent with the corresponding regimes in Fig.~\ref{fig:voids}d within the error, see the respective dashed power-law fits.

\begin{figure}[!htb]
 \centering
 \includegraphics[width=7cm]{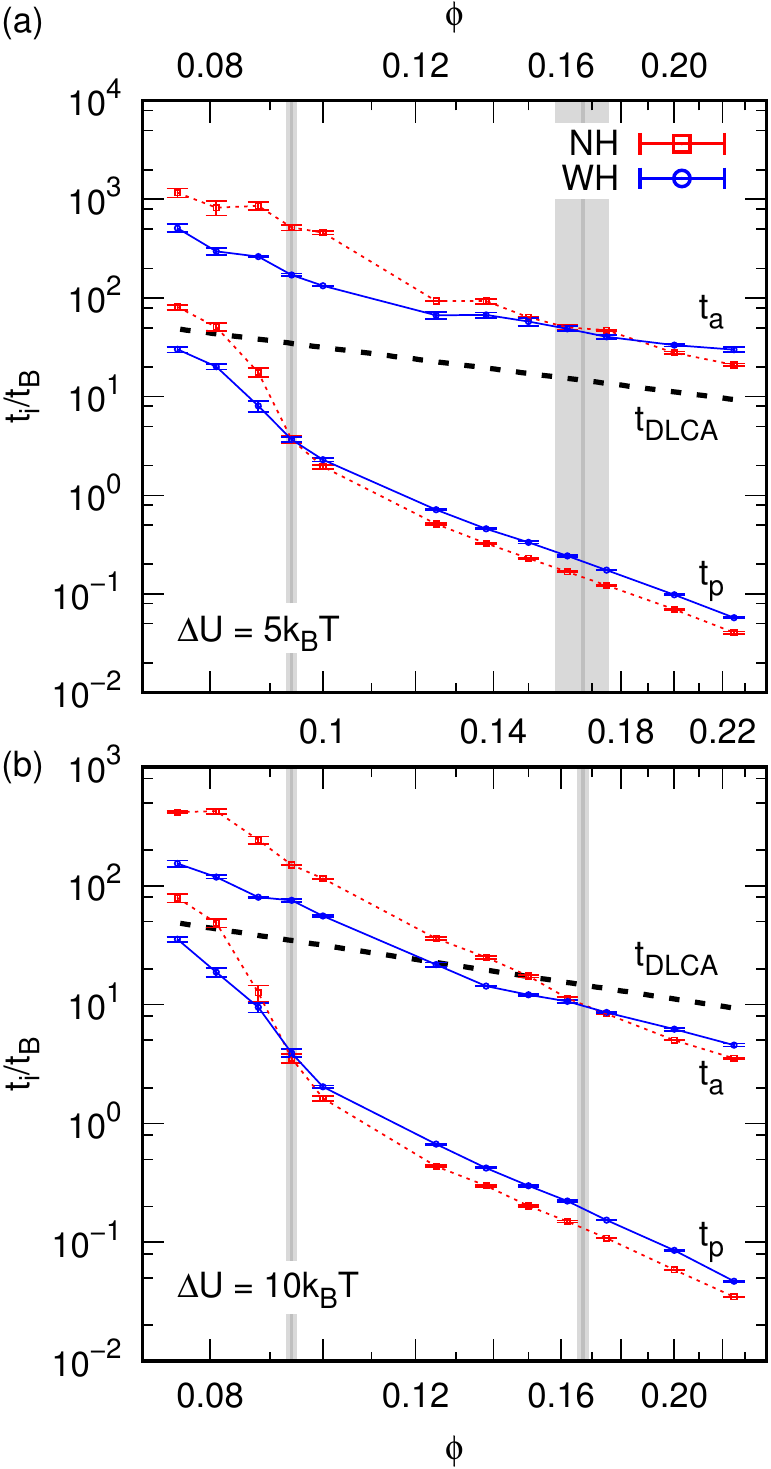}
 \caption{\label{fig:diagram}Differences in the dynamics of gelation. The times of interest, $t_{\mathrm{p}}$ and $t_{\mathrm{a}}$, for percolation and aging, respectively, as a function of the colloid volume fraction $\phi$ for (a) $\Delta U = 5 k_{\mathrm{B}}T$ and (b) $\Delta U = 10 k_{\mathrm{B}}T$. The dashed (NH) and solid (WH) curves connect the points belonging to a set of data and serve as guides to the eye. The standard error is indicated using error bars, but is typically small. The gray vertical lines indicate the $\phi$ values, for which there is a crossover, the light-gray field indicates the error therein. Finally, the thick black dashed lines show the diffusion-limited cluster aggregation (DLCA) prediction for the gel time, $t_{\mathrm{DLCA}}$, see Eq.~\eqref{eq:DLCA}, with prefactor $1$ and fractal dimension $d_{f} = 1.8$.}
\end{figure}

\subsection{Two Structural Times}

We have defined and examined two structural times: $t_{\mathrm{p}}$, when a percolating cluster first exists, and $t_{\mathrm{a}}$, where the power-law evolution of VV and $S(Q)$ show an abrupt change in exponent. We therefore classify the system's evolution into three stages: `pre-gelation' at $t < t_{\mathrm{p}}$, gelation at $t_{\mathrm{p}} < t < t_{\mathrm{a}}$, and aging at $t > t_{\mathrm{a}}$. The dependence $t_{\mathrm{p}}$ and $t_{\mathrm{a}}$ on $\phi$ is shown in Fig.~\ref{fig:diagram}. 

It is clear that HI speed up the system's dynamics below a certain $\phi$ and slow it down above this value. The crossover occurs at $\phi = 0.095 \pm 0.002$ for $t_{\mathrm{p}}$ and this effect is pronounced, while for $t_{\mathrm{a}}$ we find a crossover at $\phi = 0.16 \pm 0.01$ and $0.16 \pm 0.002$ for weak and strong gels, respectively. Interestingly, percolation occurs at the same time independent of $\Delta U$, but $t_{\mathrm{a}}$ is significantly increased upon lowering $\Delta U$, with $t_{\mathrm{a}}$ being substantially larger for systems without HI at low $\phi$.

We have also plotted in Fig.~\ref{fig:diagram} the time to form a spanning cluster in $\phi \to 0$ diffusion-limited cluster aggregation (DLCA)~\cite{witten1981diffusion} approximation:
\begin{align}
\label{eq:DLCA} t_{\mathrm{DLCA}} &\propto t_{\mathrm{B}} \phi^{-d_{\mathrm{f}}/(3-d_{\mathrm{f}})} ,
\end{align}
with a fractal dimension $d_{\mathrm{f}} \approx 1.8$~\cite{allain1995aggregation}. This behaves neither as $t_{\mathrm{p}}$ nor as $t_{\mathrm{a}}$, and does not describe our system. 

\begin{figure}[!htb]
 \centering
 \includegraphics[width=7.5cm]{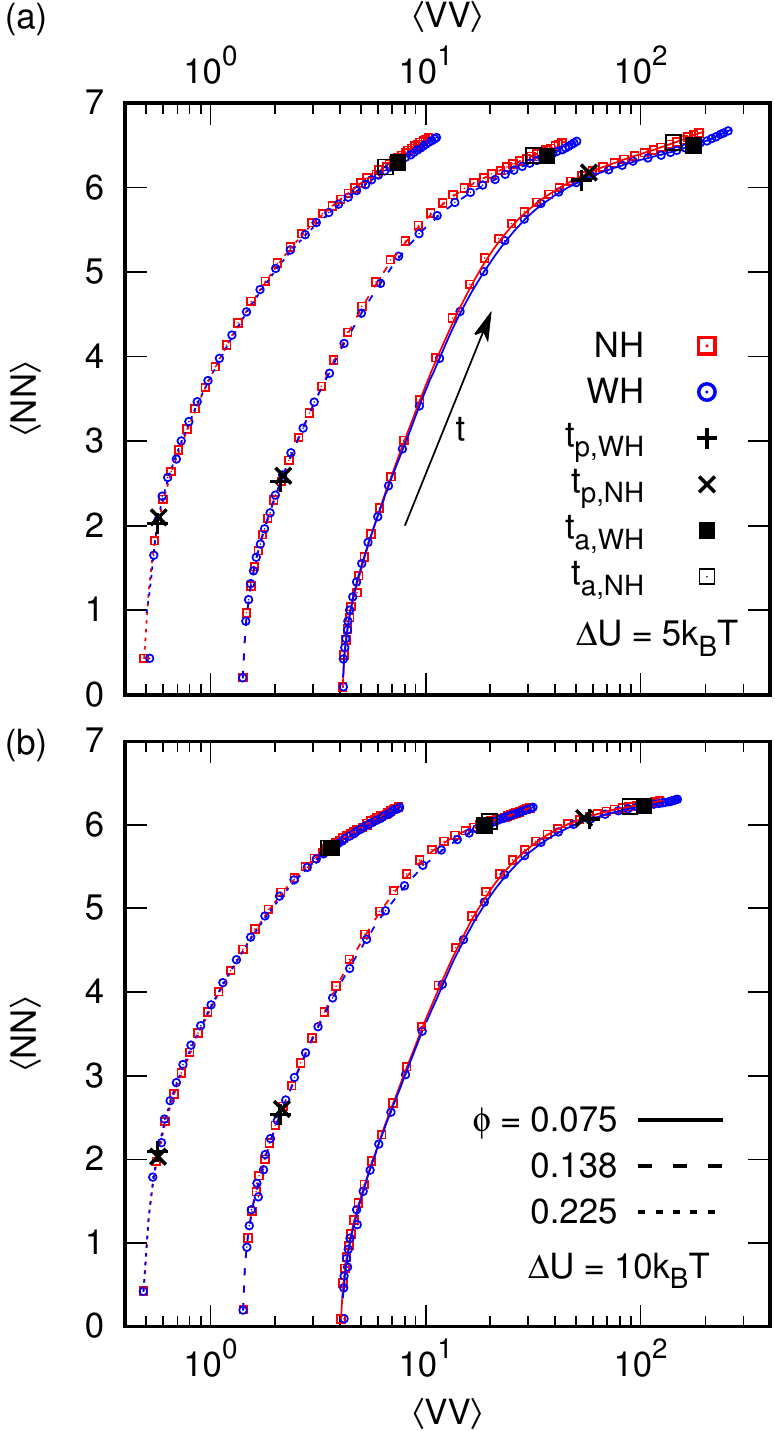}
 \caption{\label{fig:path}State-space plot for the structural evolution of colloidal gels. The space is spanned by the average void volume $\langle \mathrm{VV} \rangle$ and average number of nearest neighbors $\langle \mathrm{NN} \rangle$. The graphs show the path taken by weak, $\Delta U = 5 k_{\mathrm{B}}T$, (a) and a strong, $\Delta U = 10 k_{\mathrm{B}}T$, gels (b), respectively, for 3 different volume fractions $\phi$ as labeled; NH (red, square) and WH (blue, circle). The standard error is smaller than the symbol size in all cases and the paths are followed in time in the direction indicated by the black arrow. The symbols ($\boldsymbol{+}$,$\boldsymbol{\times}$) and ($\blacksquare$,$\square$) mark the percolation and aging time, respectively, with and without hydrodynamics.}
\end{figure}

\subsection{$\langle \mathrm{VV} \rangle$-$\langle \mathrm{NN} \rangle$ State Space Trajectories}

We have shown that HI can strongly affect the dynamics of gelation, but that when compared at two equal structural times $t_{\mathrm{p}}$ and $t_{\mathrm{a}}$, there is little difference in the particle network. This similarity can be brought out by plotting the system's trajectory in a `state space' formed by $\langle \mathrm{VV} \rangle$ and the average number of neighbors $\langle \mathrm{NN} \rangle$, Fig.~\ref{fig:path}, in which time is parametric.

$\langle \mathrm{VV} \rangle$ is mainly sensitive for the structure of the voids (and by extension the gel), while $\langle \mathrm{NN} \rangle$ probes the arrest inside the branches of the gel. The systems start with a low average number of neighbors and a low average void volume: $\langle \mathrm{VV}_{0} \rangle \approx 0.024 \phi^{2}$ holds over the entire range of considered $\phi$. As expected both parameters quickly increase as the gel forms. The path taken through state space is initially nearly insensitive to HI. After longer times, small differences between the NH and WH systems appear, predominantly for weak low-$\phi$ gels, with the latter having a slightly larger $\rm \langle VV \rangle$. For high $\phi$ and/or $\Delta U$, these differences are small or insignificant.

\section{\label{sec:discussion}Discussion}

\subsection{The Role of Percolation}

The presence of percolated or nearly-percolated structures immediately after quenching the equilibrated fluid into the spinodal region of our phase diagram is important in understanding the results presented above, and in reconciling apparently contradictory results in the literature. To see this, we first enquire whether such structures exist. Various analytical expressions in the literature, such as that based on Grimaldi's tree \textit{ansatz}~\cite{grimaldi2017tree} or from fitting the adhesive hard sphere simulations of Millar and Frenkel~\cite{miller2004phase}, are problematic for mapping onto our systems. The former ignores short-time rearrangements due to the finite-range attraction upon quenching, while using the latter involves far-fetched extrapolation (to very low `stickiness parameter'). On the other hand, recent work on attractive colloidal gels in 2D shows the importance of structural correlations for the onset of rigidity percolation~\cite{zhang2018correlated}. Moreover, early simulations of Hayward, Heerman, and Binder~\cite{hayward1987dynamic} of the kinetic Ising model give a dynamic percolation line inside the two-phase region that is almost vertical, rising from zero temperature at a concentration of $c \approx 0.16$, which translates to $\phi \approx 0.10$ (taking $c = 1$ to be random close packing at $\phi \approx 0.64$). Interestingly, the crossover in $t_{\mathrm{p}}$, Fig.~\ref{fig:diagram}, occurs at $\phi \approx 0.1$. It is therefore reasonable to suggest that there are nearly-spanning networks or spanning networks present throughout the whole $\phi$-range of our simulations, at both high and low $\Delta U$, which has three consequences.

Firstly, on the most coarse-grained level, the pre-existence of such large-scale structures is the basic physics behind the observation that, despite the large difference in dynamics, systems with and without HI follow nearly identical paths through state space, Fig.~\ref{fig:path}. Secondly, it explains the poor predictive power of Eq.~\ref{eq:DLCA}, which can only be expected to (and indeed does~\cite{whitmer2011influence}) hold in the limit of very low $\phi$, where long-range diffusion dominates, and for large $\Delta U$, where rearrangements after initial bonding are irrelevant. Finally, the crossover from the pre-existence of nearly-percolating structures to a pre-existing spanning network at $\phi \approx 0.1$ explains why we see a change in the scaling of $t_{\mathrm{p}}(\phi)$ at the same $\phi$, Fig.~\ref{fig:diagram}. Only very small local movements of these pre-existing large clusters are needed to bring about percolation. The physics here are dominated by the need to squeeze out solvent between particles that eventually bridge such clusters, so that HI slow down the dynamics, as observed. 

Once a spanning network exists, further growth of the gel structure occurs by accretion of clusters and single particles onto this backbone, depending on the volume fraction considered, until eventually the aging dynamics set in. The crossover in $t_{\mathrm{a}}$ occurs when $\phi \gtrsim 0.16$, because at lower volume fractions the system reaches the aging state through exploration of pre-arrested configurations by localized cluster movements. That is, at $\phi \approx 0.1$, collective effects are still important to reach the aging state, whereas for $\phi \gtrsim 0.16$ the squeeze-flow contributions dominate the system percolating and aging.

\subsection{Reconciling Literature Results}

We next place our results in the context of a large literature reporting simulation of colloidal gels.

First, there has been persistent reports in the literature of two different modes of gelation: equilibrium and non-equilibrium, with the former attributed to percolation \cite{Dhont1995, deKruif1990, Grant1993} and the latter to arrested phase separation \cite{Giglio1992, poon1995gelation, Verhaegh1997, manley2005glasslike, Lu2008}. There have been recent attempts to unify the two pictures by simulations~\cite{zaccarelli2007colloidal} and by experiments~\cite{helgeson2014homogeneous}, with the latter showing a crossover for percolation to arrested phase separation at $\phi \approx 0.2$ for nano-emulsions based on rheological measurements. Our results do not directly address this issue. However, they do point to the importance of whether there is a pre-existing percolating cluster at the beginning of the aggregation process.

Second, our results reconcile the apparently contradictory findings of Yamamoto~\textit{et al.}~\cite{yamamoto2008role} and Furakawa and Tanaka~\cite{furukawa2010key}. The former work reports a minor effect of HI, because their $\phi$ range is in the compaction regime. The latter, and more recently Varga~\textit{et al.}~\cite{varga2015hydrodynamics} and Royall~\textit{et al.}~\cite{royall2015probing}, report substantial structural differences at intermediate $\phi$ when gels are compared at equal times. We find the same when we compare systems at equal absolute times, rather than at equal structural times, especially when we do so for $t < t_{\mathrm{a}}$.

Third, contrary to Furakawa and Tanaka, we find gels at $\phi=0.075$ in both WH and NH systems at high and low $\Delta U$. This may reflect their protocol of comparing WH and NH systems at equal times, and/or the closeness to the gelation line of their gels. Neither do we observe a crossover from $t^{-1/3}$ to $t^{-1/2}$ in $q_{\mathrm{m}}(t)$ upon including HI. Instead, we find a weak power-law aging in the long-time regime ($t > t_{\mathrm{a}}$). Furakawa and Tanaka's finding here may reflect their limited data range. We can fit a slope of $-\sfrac{1}{3}$ to a limited range of our $q_{\mathrm{m}}$ data. While our data range is still limited in the long-time limit, our finding of slow aging agrees with Yamamoto~\textit{et al.}'s observation of a continuous gel evolution in 3D systems~\cite{yamamoto2008role}.

Fourth, we analyzed the gel structure using the three-point correlation approach introduced by Royall~\textit{et al.}~\cite{royall2015probing}. However, at equal structural times, we did not observe a difference between the local structure of systems with and without HI. The three-point correlation functions were identical within the error bar. In addition, dividing these three-point functions by the regular pair correlation function to highlight certain features, as suggested by Royall~\textit{et al.}, proved uninformative, due to rather large errors this process induced. The pronounced effect on local structure, as indicated by the authors of Ref.~\cite{royall2015probing}, could be due to their measurements having closer proximity to the gel line.

Finally, our short-time $q_{\mathrm{m}}$ data does not support the power-law dependence found by Poon~\textit{et al.}~\cite{poon1995gelation}: our limited data is consistent with a stretched exponential fit (not shown here). Clearly the mapping between the evolution of $\lambda_{\mathrm{VV}}$ and $q_{\mathrm{m}}$ breaks down at small times.

\subsection{The Gelation Boundary}

The rapid increase of $t_{\mathrm{p}}$ and $t_{\mathrm{a}}$ with decreasing $\phi$ and $\Delta U$ necessitates prohibitively long simulations to establish the effect of HI on the gelation boundary using our methodology. However, we do find that HI significantly speed up gelation in the range $0.075 \le \phi < 0.1$. This may contribute to the observed shift in the gel line reported by Varga~\textit{et al.}~\cite{varga2015hydrodynamics} and Royall~\textit{et al.}~\cite{royall2015probing}, and some of the results reported by Furakawa and Tanaka~\cite{furukawa2010key}. At low enough $\phi$, the time for cluster-cluster collision becomes longer than the time needed for single cluster compaction. Gelation fails and the system phase separates. If gravitational effects, which dominate in practice in this regime, can be eliminated, we would expect HI to shift the gel line to slightly lower $\phi$, as collective mobility should speed up the thermal exploration of a cluster's neighborhood and slow down its compaction. However, it would be very difficult to perform a matched structural age comparison, due to the divergent time scales near the gelation line. 

\section{\label{sec:conclusion} Conclusion and Outlook}

We find two regimes in our simulation of colloidal gels with and without HI. At $\phi \lesssim 0.1$, HI speed up gelation, while at $\phi \gtrsim 0.2$ HI slow down gelation, with a crossover between the two regimes at $\phi \approx 0.16$. This division appears to be in line with seemingly conflicting results reported in the literature on the effect of HI on the gel structure. There appear to be large structural effects when we compare gels with and without HI at equal times, especially for low $\phi$, but these effects are minor at high $\phi$. However, when we compare the structure at equal structural times,~\textit{i.e.}, those times at which the structures are in the same point of their structural evolution, these differences disappear and the HI effects on the structure of the gel are much smaller than the effects of parameters such as the age of the gel and the colloid volume fraction. 

The results presented here suggest several opportunities for follow-up studies. It would be worthwhile to apply this methodology to explore states closer to the gelation line and to gels with other kinds of inter-particle interactions, such as short-ranged attraction coupled with medium-ranged repulsion. Repeating our analysis using other numerical methods for hydrodynamic interactions, which make different approximations for the solution of Stokes' equations, would also be beneficial and may give closure to the discussion on the effect of hydrodynamic interactions on gelation.

\begin{acknowledgments}
We thank Marie Sk{\l}odowska-Curie Intra European Fellowship (G.A.~No.~654916) within Horizon 2020 (JdG), an EPSRC Programme Grant (EP/J007404/1) and the International Fine Particles Research Institute (WCKP) for funding; Henri Menke for \textsf{ESPResSo} support; Georg Rempfer, Paul van der Schoot, and Nick Koumakis for discussions; and Christian Holm for making his GPU cluster available.
\end{acknowledgments}

\end{document}